# Protons in the Near Lunar Wake Observed by the SARA Instrument on Board Chandrayaan-1

Y. Futaana, [1] S. Barabash, [1] M. Wieser, [1] M. Holmström, [1] A. Bhardwaj, [2] M. B. Dhanya, [2] R. Sridharan, [2] P. Wurz, [3] A. Schaufelberger, [3] K. Asamura [4]
---
Y. Futaana, Swedish Institute of Space Physics, Box 812, Kiruna, SE-98128, Sweden. (futaana@irf.se)

[1] Swedish Institute of Space Physics, Box 812, Kiruna, SE-98128, Sweden
[2] Space Physics Laboratory, Vikram Sarabhai Space Center, Trivandrum 695 022, India
[3] Physikalisches Institut, University of Bern, Sidlerstrasse 5, CH-3012 Bern, Switzerland
[4] Institute of Space and Astronautical Science, 3-1-1 Yoshinodai, Sagamihara, Japan

## Index Terms

6250 Moon
5421 Interactions with particles and fields
2780 Magnetospheric Physics: Solar wind interactions with unmagnetized bodies
7807 Space Plasma Physics: Charged particle motion and acceleration

## Abstract

Significant proton fluxes were detected in the near wake region of the Moon by an ion mass spectrometer on board Chandrayaan-1. The energy of these nightside protons is slightly higher than the energy of the solar wind protons. The protons are detected close to the lunar equatorial plane at a 140˚ solar zenith angle, i.e., ~50˚ behind the terminator at a height of 100 km. The protons come from just above the local horizon, and move along the magnetic field in the solar wind reference frame. We compared the observed proton flux with the predictions from analytical models of an electrostatic plasma expansion into a vacuum. The observed velocity was higher than the velocity predicted by analytical models by a factor of 2 to 3. The simple analytical models cannot explain the observed ion dynamics along the magnetic field in the vicinity of the Moon.

# 1. Introduction

The classical picture of the Moon-solar wind interaction is straightforward. Because the surface of the Moon is covered by non-conductive porous regolith, it behaves as a perfect absorber of the solar wind ions and electrons. The perturbations of lunar origin in the interplanetary magnetic field are extremely small, and therefore no significant effects are expected in the upstream solar wind. For example, no global scale bow shock is predicted, and consistently has never been observed. Only strongly, but locally, magnetized regions of crustal origin (often called magnetic anomalies) can interact with the solar wind under specific configurations, forming mini-magnetospheres (e.g., *Russell and Lichtenstein [1975]*; *Lin et al. [1998]; Wieser et al. [2010]*).

Because the supersonic solar wind plasma is largely absorbed by the dayside surface, a vacuum region is formed on the nightside of the Moon. The solar wind plasma refills the vacuum region, and corresponding ion signatures were observed during the wake crossings by WIND spacecraft in 1994 (*Ogilvie et al. [1996]*). One conclusion from these observations was that a large electric potential drop of ~400V close to the wake boundary has to be assumed to explain the ion signatures. Following the observations by WIND, observations of electrons and protons in the vicinity of the Moon from other missions also suggest that the significant electric potential drop does exist across the boundary (e.g., *Futaana et al. [2001]*; *Halekas et al. [2005]*; *Nishino et al. [2009a]*). A number of numerical simulations using particle-in-cell models and hybrid models of the solar wind-Moon interactions have been performed (*Farrell et al. [1998]*; *Birch and Chapman [2001]*; *Kallio [2005]*; *Trávníček et al. [2005]*), but the large potential drop has not been reproduced. *Halekas et al. [2005]* proposed a theory to explain the large potential drop by considering supra-thermal solar wind electrons.

Apart from the fluid considerations of the plasma in the vicinity of the Moon, kinetic effects have also been discussed. Using Apollo observations, *Freeman [1972]* found fluxes of light ions with mass per charge less than 10 amu/q from the zenith direction in the deep wake, but the source of these light ions was not clear. *Futaana et al. [2003]* reported non-solar wind protons with ring-like velocity distribution functions during a lunar swing-by of the Nozomi spacecraft. They interpreted that those protons are most probably reflected from a local bow shock formed in front of the lunar mini-magnetosphere, but *Holmström et al. [2010]* argued that the reflected protons from the lunar surface discovered by *Saito et al. [2008]* may also explain the Nozomi observations.

In this paper, we report on the detection of a proton flux in the deep lunar wake region by the Sub-keV Atom Reflection Analyzer (SARA) on board Chandrayaan-1. Earlier, Nishino et al. [2009a; 2009b] identified two types of proton intrusions into the wake in the ion data obtained from the Kaguya spacecraft. However, the arrival direction of proton fluxes reported in this paper is significantly different from the fluexes reported in Nishino et al. [2009a; 2009b]. The proton population discussed here is frequently seen in the SARA data, however, we selected one particular event on 25 January, 2009 when the interplanetary magnetic field and upstream condition were stable and optimal for observations, such that the interplanetary magnetic field vector was in the ecliptic plane, and perpendicular to the solar wind velocity vector.

# 2. Instrumentation

The SARA data discussed in this paper were collected on a lunar polar orbit at a height of ~100 km. SARA is composed of two sensors (*Bhardwaj et al. [2005]*; *Barabash et al. [2009]*). One of the sensors is called Chandrayaan-1 Energetic Neutrals Analyzer (CENA), which is the first-ever energetic neutral atom (ENA) sensor flown to the Moon. The other sensor is Solar WInd Monitor (SWIM), an ion mass analyzer to monitor the solar wind and to study the ion environment around the Moon. Only data obtained by the SWIM sensor are used in this study.

The SWIM sensor is a compact electrostatic ion mass analyzer with a fan-shaped aperture (~7°×160°) and an angular resolution of ~7°×10° (depending on the looking direction). The number of angular pixels is 16 (maximum). For this study SWIM was operated with an energy range of ~100 to 3000 eV/q covered by 16 logarithmically separated energy-per-charge bins. SWIM also has a moderate mass resolution of $m/\Delta m \sim 2$ (*McCann et al. [2007]*; *Barabash et al. [2009]*).

Figure 1 shows the SWIM and CENA fields of views (FoVs). The SWIM bore sight is along the $+z_{sc}$-axis, and the aperture plane is perpendicular to the $y_{sc}$-axis. Here the subscript SC denotes the spacecraft reference frame. The nominal spacecraft attitude is the nadir pointing. During nadir pointing, the $+x_{sc}$ axis always points toward the lunar surface. The velocity vector of the spacecraft is nominally along either the $+y_{sc}$ or the $-y_{sc}$-axis. During the period of the observation discussed in this paper, the $-y_{sc}$-axis was co-aligned with the velocity vector. As seen in Figure 1, the SWIM aperture is perpendicular to the spacecraft velocity vector, and some of the SWIM angular sensors (CH-0 to -5) point toward the lunar surface. This configuration means that Chandrayaan-1 can be considered as a spinning satellite revolving around the $z_{sc}$-axis with a rotation period equal to the spacecraft orbit around the Moon (~118 min). The SWIM FoV plane (the half-plane of $\pm x_{sc}$ and $+z_{sc}$) can cover ~2π angular space (half hemisphere) in half of the orbital period (~59 min). The three-dimensional (3-D) velocity distribution function of the solar wind can be measured only when the spacecraft is at the dayside and the SWIM FoV is close to the ecliptic plane. Therefore, the solar wind can be observed only once per orbit when the spacecraft is close to the lunar dayside equator. The 3-D velocity distribution of the protons allows us to calculate the density. We calculated the density of the protons by numerical integration over the observed flux.

In this paper, the lunar-centric solar ecliptic (LSE) coordinate system is used. The Moon-Sun line is the $+x$ axis, the velocity vector of the Sun motion relative to the Moon is the $+y$ axis, and the $+z$ axis completes the right-handed system. Indeed, the differences of the axis directions with the geocentric solar ecliptic (GSE) frame (+*x* axis is the Earth-Sun line, +*z* axis is normal to the mean ecliptic plane of the Earth pointing to north, and +*y* axis completes the right-handed system) is in general very small (<0.3˚), and it was ~0.18˚ for the day of the observation.

# 3. Observation

Figure 2 shows the energy-time spectrogram of ion counts observed by SWIM on 25 January 2009. The lunar phase was 2 days before the new Moon. Because the Moon was in the undisturbed solar wind, no effects from the Earth's bow shock or magnetosphere were expected. Figure 3 shows the Chandrayaan-1 orbit corresponding to the SWIM observations discussed here (orbit 942). The motion of the spacecraft was from north to south on the dayside, and from south to north on the nightside. The Sun aspect angle, the angle between the spacecraft orbital plane and Sun-Moon line, was ~40˚.

In the SWIM data shown in Figure 2, five distinct populations of ions, labeled as (A)-(E), can be clearly identified. The strongest flux is from the solar wind protons with an energy of ~500-600 eV/q (population A). The corresponding solar wind velocity is ~310-340 km/s. At the same time, we can see the population (B) at a slightly higher energy range (*E/q*~1 keV/q) that consists of the α-particles from the solar wind. Because the α-particles are doubly charged, the actual energy is ~2 keV, and the velocity was similar to that of the solar wind protons (A). These two populations are observed on the lunar dayside close to the lunar equator from directions 11 and 12, as expected.

The ion populations (C) and (D) are also detected on the dayside. Both populations are composed of protons as established by mass analysis (not shown here). The population (C) comes from the surface and has a broadened energy spectrum. These ions are backscattered protons from the lunar surface similar to the observations reported by *Saito et al. [2008]*. Ions of population (D) are the backscattered protons accelerated by the convection electric field of the ambient solar wind electric field, which was also suggested by *Saito et al. [2008]*. Note that the accelerated protons (D) are absent on the second orbit. The signatures of this population will be discussed in the Discussion section.

The population (E) is a faint proton flux that can be seen in the deep nightside region. The energy of the population (E) is 0.5 to 1 keV, which is slightly higher than that of the solar wind. The densities of the solar wind (A) and the population (E) can be calculated by the integration of the observed flux. The density is 1.7 $cm^{-3}$ for the solar wind (A) and $(3\text{-}4)\times10^{-3}$ $cm^{-3}$ for the population (E). Notably, the density calculation is not straightforward especially for the case of low ion flux, and therefore a large ambiguity may be included. In addition, the solar wind density is an underestimation because anomalously lower efficiencies than the ground calibration were found in the solar wind channels of the sensor. This is consistent with the solar wind density of 6-8/cc obtained from WIND/SWE instrument during this period. On the other hand, no such lower efficiencies were found in the channels where the population (E). The density ratio between the night side ions and the solar wind is then $(0.5\text{-}2)\times10^{-3}$.As mentioned in the Instrumentation section, SWIM can measure 3-D distribution functions with the help of the spacecraft motion and its pointing. In addition, because the ion populations (A) and (E) are both narrow beams with thermal extents of ~10° (FWHM), as estimated from the velocity spread, almost all of these beams can even be measured at a specific time.

ACE magnetic field data corrected for the propagation time to the Moon were used in this study to understand the interplanetary magnetic field (IMF) condition at the Moon. The separation between ACE and the Moon was ~180 $R_E$ during the observations, where $R_E$ is the Earth radius (6378 km). Considering the velocity of the solar wind of 310 to 340 km/s as observed by SWIM, the solar wind propagation between ACE and the Moon was 56 to 61 minutes. Hence, we have assumed that the time difference from ACE to the Moon was 1 hour. Figure 4 shows the interplanetary magnetic field data observed by ACE in the GSE coordinate system. The difference between the GSE and the LSE frame is small (~0.18°) enough to consider them as identical. The ACE data obtained between 12:00-18:00 UT (lower axis) were shifted to 13:00-19:00 UT at the Moon (upper axis). The magnitude of the IMF was stable at 3-4 nT over the period of interest. The magnetic field vector elevation angle, $\sin^{-1}(Bz/B)$, was almost zero, which means that the magnetic field vector was closely confined to the $x_{\mathrm{LSE}}$-$y_{\mathrm{LSE}}$ plane during the observations. The azimuthal angle, $\tan^{-1}(By/Bx)$, was 135° before 13:00 UT (~14:00 UT at Moon) following the Parker spiral. Between 13:00 and 14:00 UT (14:00 and 15:00 UT at Moon), a slightly fluctuating IMF azimuthal angle was observed. Afterward, the angle changed to 90°, meaning that the IMF direction was almost perpendicular to the solar wind velocity during the time the proton population (E) was observed.

# 4. Discussion

The IMF configuration and the change of its direction can consistently explain the characteristics of the accelerated protons (D). Because the direction of the convective electric field ($-\boldsymbol{v}\times\boldsymbol{B}$) points toward the northern hemisphere throughout the observations, it is natural that the accelerated protons (D) are only detected in the northern hemisphere (Figure 2). The convective electric field is $\boldsymbol{E}=-\boldsymbol{v}\times\boldsymbol{B}=$ ~300 km/s × 3 nT × cos 45° = ~0.6 mV/m = ~1.1 kV/$R_M$ where $R_M$ is the lunar radius (1738 km). The estimated energy is consistent with the observed energy of the accelerated protons. The change in the IMF direction at 15:00 UT at the Moon can explain the disappearance of the proton flux (D) on the second orbit. For the first orbit, the azimuthal angle of the magnetic field ($\tan^{-1}(By/Bx)$ ) was ~135° to the Sun-Moon line. This angle means that the magnetic field direction was almost in the same plane as the orbital plane of Chandrayaan-1 (Figure 5). The backscattered protons are accelerated by the electric field, and then start gyrating around the magnetic field (Holmström et al., 2010). This $\boldsymbol{E}\times\boldsymbol{B}$-drift motion is confined in the plane perpendicular to the IMF and the orbital plane in this case. Because the SWIM FoVs are perpendicular to the orbital plane, the ion gyration plane is favorable to be observed by SWIM. After the magnetic field direction changed to 90° prior the second orbit, the ***E***×***B***-drift motion was confined in the *x*-*z* plane. Because the ambient convection electric field accelerates protons in the +*z* direction quickly, the *z*-component of the velocity vector of the gyrating proton dominates over its *x*-component. Such particles cannot be observed by SWIM because they do not enter the SWIM FoV, which is oblique to the motion of these particles.

The ion flux on the night side (E) is not simple to interpret. As was mentioned in the Introduction, there were nightside ion observations by the Apollo lander reported by *Freeman [1972]*, however, more detailed investigations could not be conducted because of the uncertainty of the upstream solar wind conditions. WIND (*Ogilvie et al. [1996]*; *Mall et al. [1998]*) and NOZOMI (*Futaana et al. [2003]*) reported lunar-related ions, but these observations were not conducted in the near lunar wake.

Recently, the analysis of the Kaguya data obtained at a height of 100 km identified two mechanisms of the intrusion of protons into the near lunar wake. The first mechanism is the acceleration of the solar wind protons into the lunar wake by an electric potential at the wake boundary (*Nishino et al. [2009a]).* The intrusion takes place at the wake boundary region where the solar wind velocity vector is perpendicular to the interplanetary magnetic field, with an asymmetry depending on the Larmor motion of the solar wind protons affected by the large inward electric field (~400 V). The second mechanism is the transport of the backscattered protons from the dayside by the ***E***×***B***-drift (*Nishino et al. [2009b]*). This intrusion can be realized when the gyro-radius of backscattered protons is of the same order as the lunar radius. Indeed, the gyro-radii of protons is ~$1.4\times10^3$ km for zero initial velocity under a solar wind velocity of 400 km/s and magnetic field of 3 nT.

However, these theories cannot be applied to explain the night side ion flux observed by SWIM, which was propagating along the magnetic field. Figure 6 shows the observed velocity distribution functions for the solar wind (a) sliced in the ecliptic plane and (b) along the direction perpendicular to the ecliptic plane ($v_z$). The fan-shaped filled pseudo-color image shows the phase space density of ions observed on the nightside (14:54-15:00 UT), and the contour lines are the solar wind protons and α-particles (13:57-14:03 UT). The magnetic field in the solar wind reference frame estimated from the ACE data is also superimposed on the plot. The cross mark indicates the solar wind velocity vector from the WIND/SWE data. A relatively large y-component velocity of the solar wind in SWIM data is probably from instrument effects. The channels where the main solar wind component are expected have in general lower efficiency than in the ground calibration. The reasons are yet unknown. There are no such problems in the channels where the nightside flux was observed. It is clear from Figure 6 (a) that the observed flux is along the direction of the interplanetary magnetic field, which is different from the Kaguya measurements. The theories to explain the Kaguya measurements then do not apply. Additionally, it is noted that Figure 6 (b) shows that we can measure the 3-D velocity distribution function with the assistance of the spacecraft motion and the nadir pointing.

A very simple 1-D model, which is based on the classical theory of a plasma expansion into vacuum along the magnetic field line, has been employed to explain the plasma distribution in the lunar wake using an analytical formulation [*Ogilvie et al. [1996]*; *Halekas et al. [2005]*]. Even though such a 1-D model is too simple for detailed discussions on the physics in the lunar wake, we use this model to explain the origin of the observed nightside protons (E). Because the solar wind plasma (both protons and electrons) has higher mobility in the direction parallel to the magnetic field line than in the perpendicular directions, the solar wind plasma past the terminator immediately starts filling the lunar wake. In the solar wind rest reference frame, the theory of 1-D expansion into vacuum can be applied to the expansion into the lunar wake. The configuration of the IMF direction perpendicular to the solar wind velocity vector on 25 January 2009 is favorable for applying the 1-D expansion theory. Because the pressure gradient is parallel to the magnetic field direction, the diamagnetic current ($-\nabla p \times \mathbf{B}$) can be neglected.

The gyromotion in the direction perpendicular to the magnetic field arising from the thermal speed may play a role because the gyroradius is ~100 km (for protons assuming a thermal velocity of 30 km and a magnetic field strength of 3 nT), which is comparable to the spacecraft altitude. However, the gyromotion is less significant compared to the parallel expansion, and we can still apply the 1-D approximation to this event. The reason is because the gyroperiod (~22 s) is long enough compared to the travel time of ~3 s for the nightside protons from the terminator to the observation point, and thus the protons experience only a part of the gyration until they are observed and the trajectory of the protons in the Moon frame does not change significantly.

The 1-D formulation of a plasma expansion into a vacuum is summarized by *Samir et al. [1983]*. The following assumptions are made: a) the electrons are always in equilibrium with the electrostatic potential employed by Boltzmann's relations (e.g., *Crow et al. [1975]*); b) the solar wind electrons follow the Maxwell distribution functions with a constant temperature, $T_e$, over the system; c) the ion temperature, $T_i$, is zero; and d) charge neutrality. These four assumptions are rather realistic, but the problematic assumptions that have been introduced for the study of the plasma expansion into the lunar wake are as follows: e) we neglect of the surface potential; and f) we ignore the surface absorption. The last two assumptions are quite difficult to include in analytical models, and therefore, they have been assumed by many authors explicitly or implicitly.

The plasma parameters, the ion density ($n_i$), ion velocity ($v_i$), and the electric potential ($V$) can be described as a function of the distance from the vacuum boundary at the initial state, $s$. The set of the equations could be formulated as:

$$n_i = n_e = n_0 \exp\left(-\frac{s}{V_{ia}t} - 1\right) \qquad (1)$$

$$v = \frac{s}{t} + V_{ia} \qquad (2)$$

$$V = -\frac{kT_e}{e}\left(\frac{s}{V_{ia}t} + 1\right) \qquad (3)$$

where $t$ is time, $n_0$ is the undisturbed plasma density, $n_e$ is the electron density, $e$ is the elementary charge, $M$ is the mass of the proton, and $k$ is the Boltzmann constant. $V_{ia}=(kT_e/M)^{1/2}$ is the ion acoustic velocity under the assumption of c) $T_i$ =0. WIND [Ogilivie et al., 1998] and Lunar Prospector [Halekas et al., 2005] observations can be better explained by assuming the solar wind electron velocity distribution by a κ-distribution than by a Maxwell distribution. In particular Lunar Prospector observations are conducted at an altitude similar to our observation, and the electron distributions are consistent with a model using a κ-distribution [Halekas et al., 2005]. Therefore, comparison with the κ-distribution is also worthwhile to understand the ion dynamics in the lunar wake.

Table 1 shows the results of the calculated density and velocity at the spacecraft position. The SARA observations show a 2 to 3 times higher velocity than the model calculations. The observed density is lower than the model calculations by a factor of 2 to 25. However, the density ratio calculated by the 1-D models depends quite sensitively on the solar wind electron temperature, which we do not know for this observation. In addition, the density calculation from the electrostatic analyzer data is not straightforward. Therefore, we can only say that there is a possibility that the observed density ratio is lower than that given by the models. On the other hand, the velocity measurement by the electrostatic analyzer is more reliable, and the dependence on the models is quite small. Therefore, we can conclude that the observed velocity is significantly different from the model. This conclusion may contradict the results of the electron distribution by the Lunar Prospector [*Halekas et al., 2005*]. The contradiction between the electron distribution that is consistent with 1-D model and the observed velocity of the protons higher than the same 1-D model is yet an open question, but it should be investigated in the future.

The reason for the higher velocity (2 to 3 times) of the observed protons is also an open question. One possible reason is that the 1-D model is too simple to reproduce the lunar wake plasma physics quantitatively. The ion absorption by the lunar surface, which was neglected in the models, may potentially explain the higher velocity (and possibly also the lower density) of the nightside ions. As soon as the solar wind electrons or ions are absorbed by the surface, the self-similar solutions cannot be used anymore. The theoretical estimate from equation (1) is that $e^{-1}$ (~36%) of the solar wind ions are absorbed by the lunar surface (see Appendix A). This large absorption rate may explain the possible lower proton density in the observations. In addition, the extra acceleration may be explained by the selection effect: only protons with a high velocity component along the magnetic field can reach the observation point.

The surface potential at the terminator region may also play a role [*Kimura and Nakagawa, 2008*], particularly if one considers the plasma absorption at the lunar surface. Due to the high speed of the electrons, the solar wind electrons are absorbed by the lunar surface at the terminator and at the nightside hemisphere of the Moon. Due to the low conductivity of the lunar regolith, the absorbed electrons are "attached" at the lunar surface, generating the negative surface potential until the equilibrium of the influx of solar wind electrons and protons is satisfied.

*Kimura and Nakagawa, 2008* conducted a 2-D particle simulation to investigate the effect of the surface potential at the terminator. They claim that at the terminator, the electric potential becomes 60-80V negative due to the electron attachment to the lunar surface. The potential drop may help accelerate the protons into the wake as observed by SWIM. When they removed the surface charging effect from their model, the acceleration of the ions decreases less at 6.5 $R_L$. The effect of the negative electric potential in the terminator region by the electron attachment may also contribute to the accelerated proton signatures observed by SWIM. However, note that *Kimura and Nakagawa, [2008]* uesd an unrealistically large Debye length (at most $R_L/8$), and therefore, a direct comparison with the data from SARA (100 km altitude) is quite difficult. Detailed comparison with simulation results using more realistic parameters are needed to understand the ion dynamics in the wake close to the Moon.

## 5. Summary

We analyzed data from the ion spectrometer SWIM on board the Chandrayaan-1 spacecraft on 25 January 2009. During the observations, the IMF conditions were stable and the geometry of the upstream electromagnetic field was relatively simple.

Three ion populations in addition to the nominal solar wind ions (both protons and α-particles) are identified in the SWIM data. On the dayside, backscattered protons and accelerated backscattered protons are observed, and they are similar to the populations observed earlier by Kaguya (*Saito et al. [2008]*). These observations can be explained by single particle motions in the interplanetary magnetic field and the convective electric field.

We also detected proton fluxes in the lunar wake region. The observed position was ~50° from the terminator inside the lunar near wake at a height of 100 km. The flux propagates along the magnetic field in the solar wind frame; therefore, the gyro-motion, interplanetary magnetic field, and the convective electric field cannot play a role. The proton energy was ~700 eV, which was slightly higher than the solar wind bulk energy of ~550 eV during the observation period.

The prediction of the 1-D models could not explain the velocity of the observed protons as it was 2 to 3 times higher than the velocity given by the model. The observed velocity is higher than the prediction by the models. The reason of the difference in the velocity is yet an open question, but the surface absorption effect, which is neglected in the analytical models, and the negative surface potential of the Moon at the terminator region and the nightside surface may be one possible reason. The absorption of the plasma particles and the resulting large electric potential at the lunar surface could be significant for understanding the kinetics of the solar wind ions in the low-altitude wake of the Moon.

# Appendix A. The absorption ratio of protons at the lunar surface

Here we calculate the absorption ratio of the solar wind protons at the lunar surface close to the terminator. The coordinate system used in this appendix is drawn in Figure A1. When the solar wind protons expand into the lunar wake, a rarefaction wave is formed. The rarefaction wave front, $s_{\mathrm{wf}}$, is the boundary separating the undisturbed from the disturbed solar wind. From the equation (1), we know that the wave front propagates with a velocity of the ion acoustic speed:

$$s_{\mathrm{wf}} = -V_{ia} t \qquad \text{(A1)}$$

The total amount of solar wind plasma that is affected by the vacuum expansion, *Nt*, is an integration of the density as

$$N_{\mathrm{t}} = \int_{s_{\mathrm{wf}}}^{\infty} n_i \mathrm{d}s = n_0 V_{ia} t \quad \text{(A2)}$$

Conversely, the location of the lunar surface, $s_{\mathrm{ls}}$, is purely geometric and can be described as

$$s_{\mathrm{ls}} = R_{\mathrm{M}} - \sqrt{R_{\mathrm{M}}{}^2 - x^2} = R_{\mathrm{M}} - \sqrt{R_{\mathrm{M}}{}^2 - (v_{\mathrm{sw}} t)^2} \qquad \text{(A3)}$$

The plasma that passes through the lunar surface (in reality, the plasma is absorbed), Na, is again an integration of the density as follows:

$$N_{\mathrm{a}} = \int_{s_{\mathrm{ls}}}^{\infty} n_i \mathrm{d}s \qquad \text{(A4)}$$

Substituting (5) and (A3) into (A4), the absorbed plasma density is obtained as

$$N_a = n_0 V_{ia} t \exp\left( -\frac{s_{ls}}{V_{ia} t} - 1 \right) \qquad \text{(A5)}$$

The absorption rate of the solar wind protons at the lunar surface can be calculated as *Na/Nt*. This formulation is only valid just after the vacuum expansion starts because the surface absorption violates the self-similar solution of (1). Therefore, one must take the limit of the time *t* to 0, which results in $Na/Nt \rightarrow e^{-1}$ ($t \rightarrow 0$).

# Acknowledgment

We thank the ACE MAG instrument team and the ACE Science Center for providing the ACE data. We also thank the WIND/SWE instrument team for the provision of the solar wind velocity and the density data. The efforts at the Swedish Institute of Space Physics were supported in part by European Space Agency (ESA). The effort at the University of Bern was supported in part by ESA and by the Swiss Science Foundation. The efforts at the Space Physics Laboratory, Vikram Sarabhai Space Centre were supported by Indian Space Research Organization (ISRO).

# Figures

**Figure 1:**

The CENA and SWIM apertures and the numbering of the viewing directions relative to the spacecraft (SC) reference frame. The SWIM has a ~7×160˚ aperture divided into 16 viewing directions. The CENA aperture is 10×160˚ divided into seven azimuthal channels. In the nominal nadir pointing, the spacecraft velocity is parallel to the $+y_{sc}$ or $-y_{sc}$ axis, and the $+x_{sc}$ axis points to the lunar center.

**Figure 2:**

The energy time spectrograms observed on 25 January 2009 over two consecutive orbits 942 and 943. From top to bottom, the energy-time diagrams for the observed ion counts coming from the surface (below local horizon), limb (toward the horizon), and space (above the horizon) are shown. At the top, the time intervals when the spacecraft was in the lunar shadow (eclipse) as well as the equator crossings are indicated. The spacecraft location (north or south hemisphere) is also indicated. Five distinct ion populations are identified, and are labeled from A through E.

**Figure 3:**

The Chandrayaan-1 orbit in the LSE coordinate system (the $+x$ axis is the Moon-Sun line, the $+y$ axis the velocity vector of the Sun motion relative to the Moon, and the $+z$ axis completes the right-handed system) between 13:30-15:28 UT on 25 January 2009. The orbital period was ~118 minutes.

**Figure 4:**

Interplanetary magnetic field data observed by ACE between 12-18 UT on 25 January 2009. Considering the propagation time from ACE to the Moon of 1 hour, the solar wind at the Moon corresponds to 13-19 UT (upper axis). From top to bottom: the magnitude, the latitudinal component, and the longitudinal component in the GSE frame of the magnetic field are displayed.

**Figure 5**

Illustration of the proton transport by the gyromotion under the two different upstream magnetic field directions (a)135° and (b) 90°. The illustration is seen from the north pole (i.e., projection to the *x-y* plane). The SWIM FoV is drawn at the north pole for simplicity. At the north pole, the acceptance is only 10° degrees under the nadir pointing. (a) If the IMF direction is 135°, the IMF is almost in the same plane as the orbital plane of the observation. Therefore, the ExB drift is perpendicular to the orbital plane, and SWIM may be able to see the gyrated protons. (b) Conversely, when the IMF direction becomes 90°, the gyration is only in the *x-z* plane; therefore, SWIM cannot detect the gyrated protons at all.

**Figure 6:**

(a) Velocity distribution functions of the solar wind (contour lines) and the nightside ions (filled polygons) in the lunar ecliptic plane. The velocity distribution functions of solar wind ions and the nightside ions are the average of the observation between 13:57-14:03 and 14:54-15:00 UT, respectively. During the period, the field of view of SWIM is in the ecliptic plane. The velocity of the solar wind from WIND/SWE data is shown by the cross. The magnetic field line is superimposed in the solar wind reference frame. (b) The solar wind velocity distribution as a function of Vz. The Vz component can be measured by the assist of the spacecraft motion and its nadir pointing. The data were taken between 13:55 and 14:05 UT.

**Figure A1.**

The coordinate system used for the calculation of the absorption rate. The *x*-axis is the Moon-Sun line identical to the LSE frame. The *s*-axis is perpendicular to the *x*-axis along the magnetic field, but its origin is the wake boundary at the terminator. The rarefaction wave front and the lunar surface are drawn by dashed lines.

Table 1: Summary of the calculation. *1 Taken from Newbury et al., 1998. *2 Taken from Halekas et al., 2005.

| | Velocity | Density (ratio) |
|---|---|---|
| SARA observation | 300-400 km/s | 0.05-0.2% |
| Model by Maxwellian electrons<br>[Samir et al. (1983)]<br>Te=141000±38000 K[*1] | 161-170 km/s | 0.4-1.2% |
| Model by κ-distribution<br>[Halekas et al. (2005)]<br>Te=141000K[*1]; κ=4.5[*2] | 185 km/s | 0.9% |

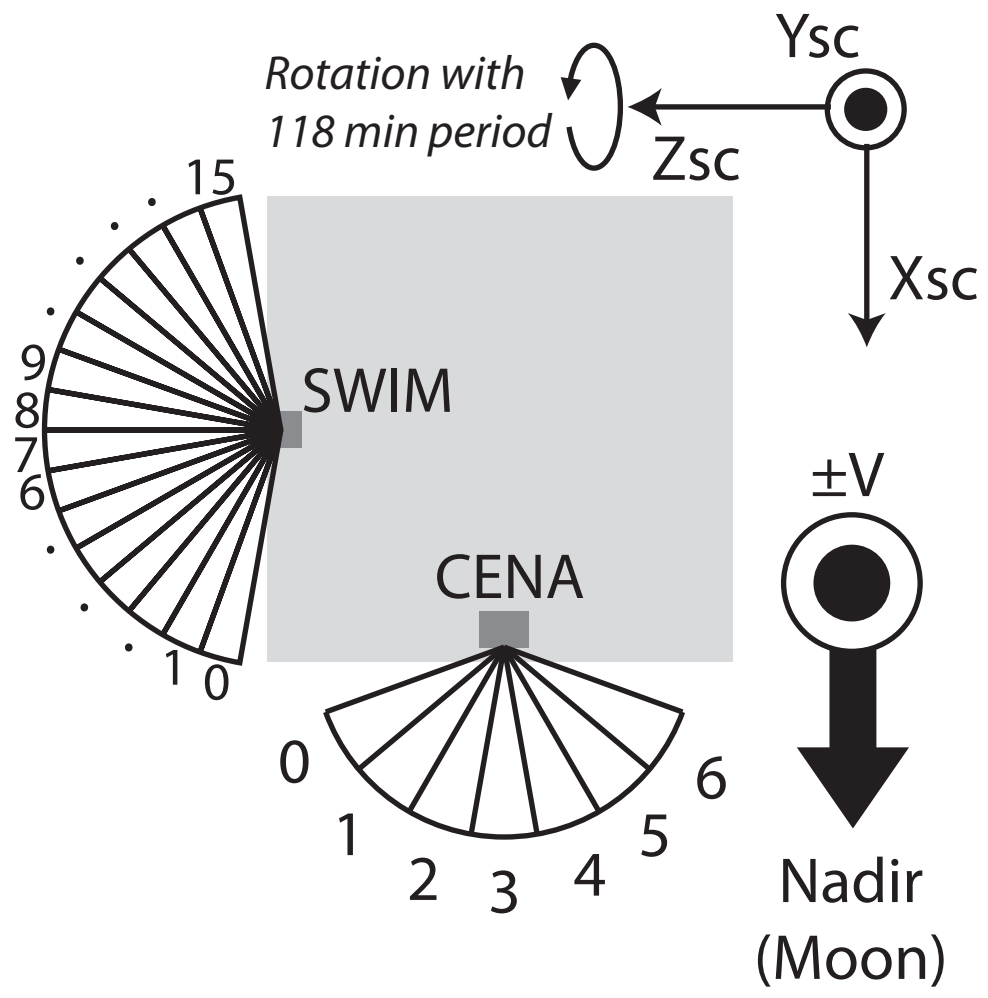

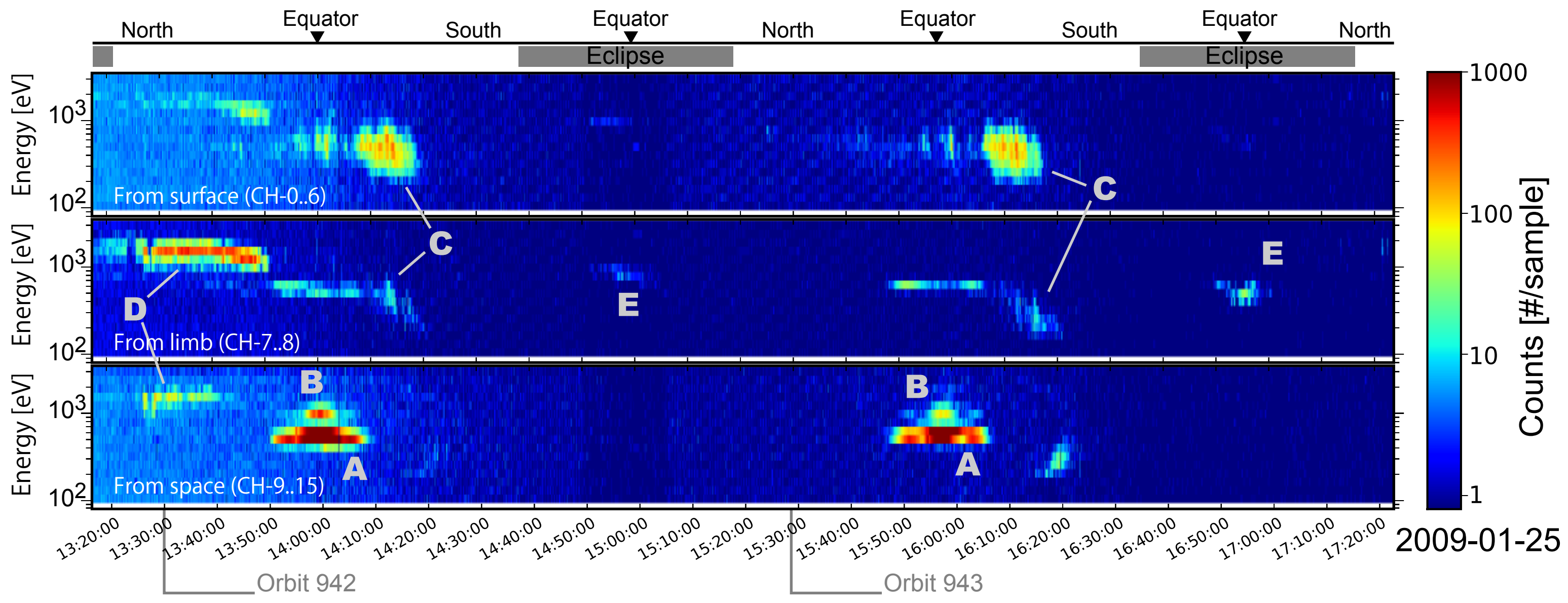

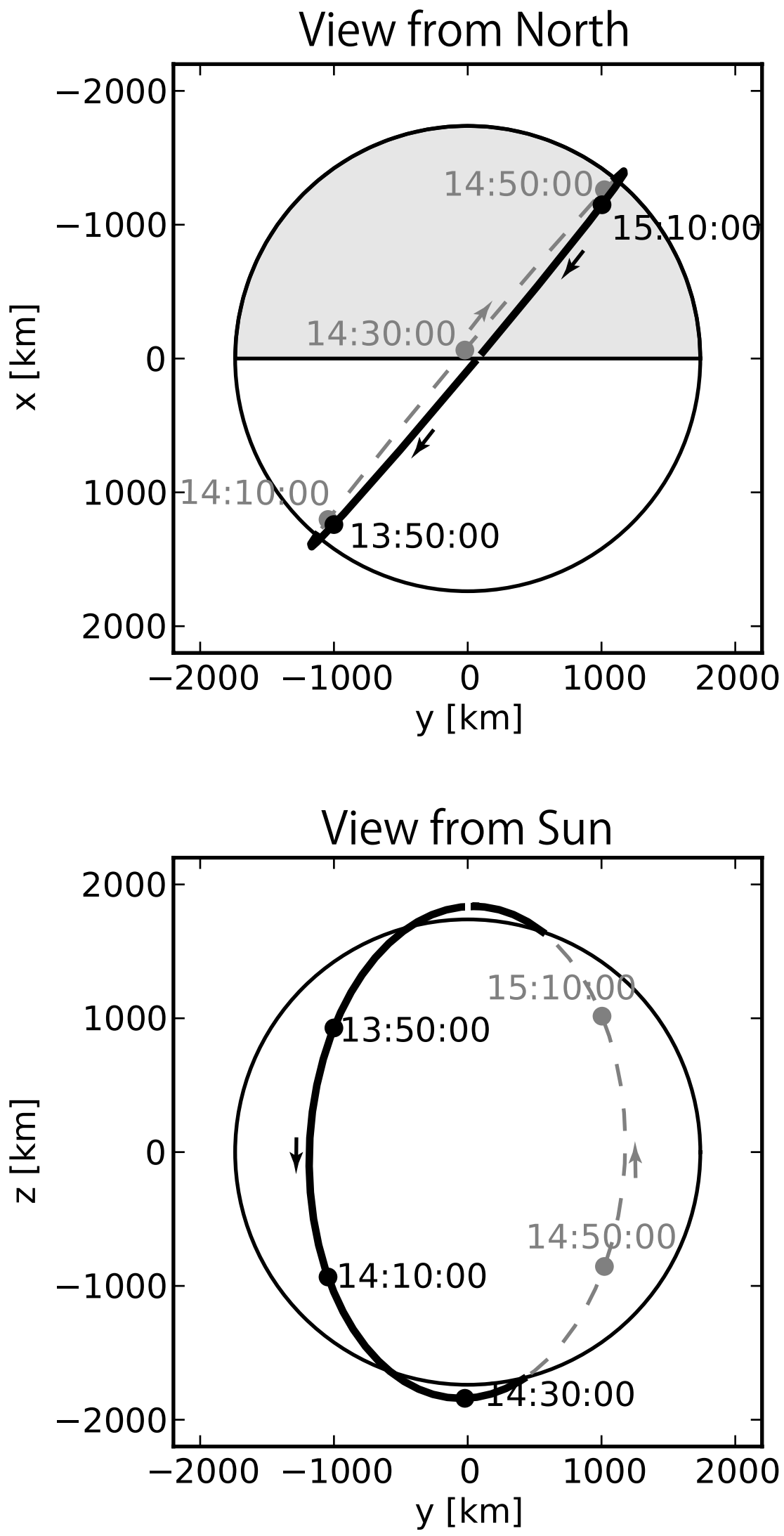

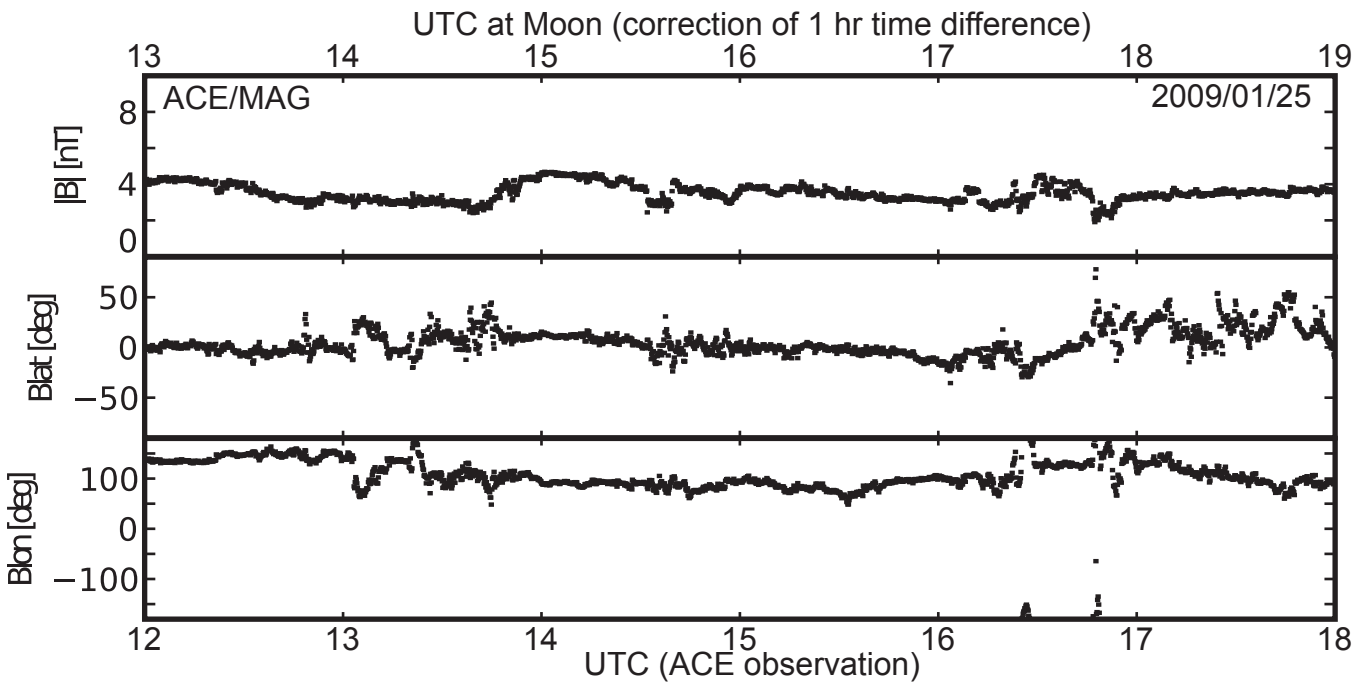
UTC at Moon (correction of 1 hr time difference)
13
14
15
16
17
18
19
ACE/MAG
2009/01/25
|B| [nT]
8
4
0
Blat [deg]
50
0
−50
Blon [deg]
100
0
−100
12
13
14
15
16
17
18
UTC (ACE observation)

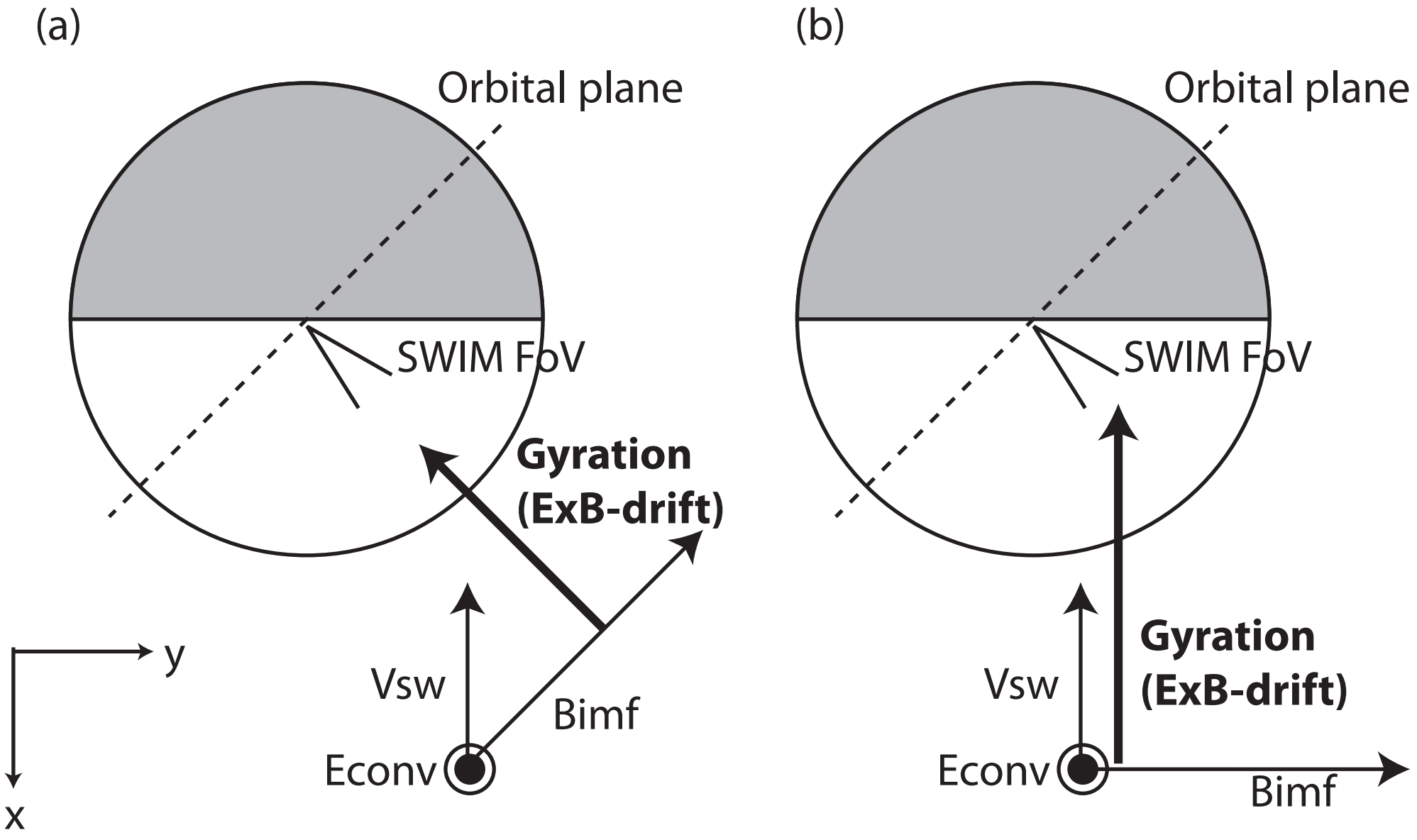

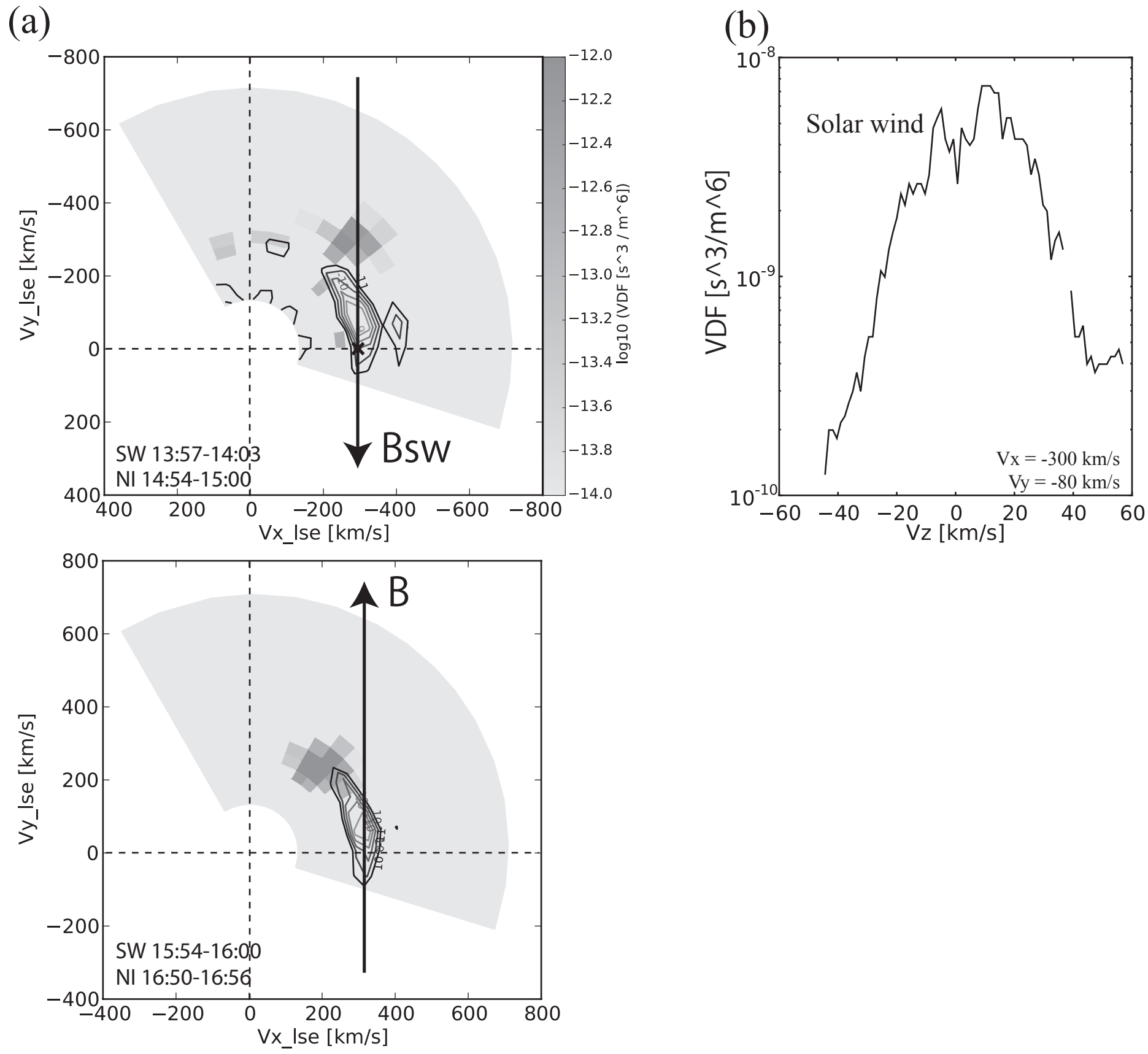

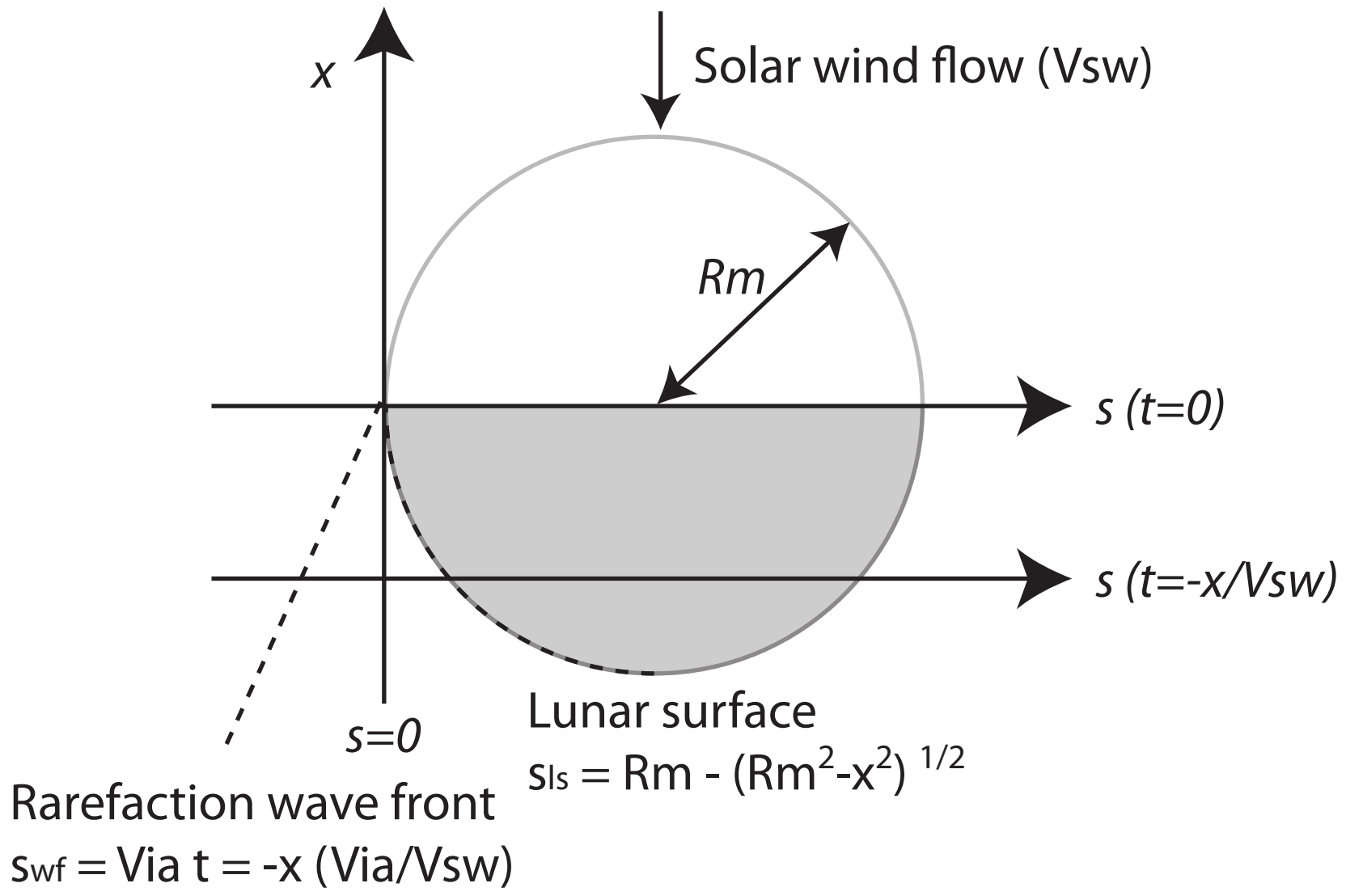
x
Solar wind flow (Vsw)
Rm
s (t=0)
s (t=-x/Vsw)
s=0
Lunar surface
$s_{ls}$ = Rm - (Rm$^2$-x$^2$) $^{1/2}$
Rarefaction wave front
$s_{wf}$ = Via t = -x (Via/Vsw)